\documentclass{sigchi-ext}
\usepackage[T1]{fontenc}
\usepackage{textcomp}
\usepackage[scaled=.92]{helvet} 
\usepackage{graphicx} 
\usepackage{balance}  
\usepackage{booktabs} 
\usepackage{ccicons}  
\usepackage{ragged2e} 
\usepackage[table]{xcolor}
\usepackage{array}
\newcolumntype{P}[1]{>{\centering\arraybackslash}p{#1}}
\usepackage{colortbl}
\usepackage{caption}
\usepackage{subcaption}
\usepackage[official]{eurosym}

\copyrightinfo{\scriptsize Permission to make digital or hard copies of part or all of this work for personal or classroom use is granted without fee provided that copies are not made or distributed for profit or commercial advantage and that copies bear this notice and the full citation on the first page. Copyrights for third-party components of this work must be honored. For all other uses, contact the owner/author(s). Copyright is held by the author/owner(s). \\
{\emph{CHI'16 Extended Abstracts}}, May 7--12, 2016, San Jose, CA, USA. \\
ACM 978-1-4503-4082-3/16/05. \\
\href{http://dx.doi.org/10.1145/2851581.2892318}{http://dx.doi.org/10.1145/2851581.2892318}}
\title{VapeTracker: Tracking Vapor Consumption to Help E-cigarette Users Quit}

\numberofauthors{6}
\author{\vspace{-1pc}
    \alignauthor{%
    \textbf{Abdallah El Ali}\\
    \affaddr{Media Informatics and Multimedia Systems} \\
    \affaddr{University of Oldenburg, Germany} \\    
    \affaddr{abdallah.el.ali@uni- oldenburg.de \bigskip} }\alignauthor{%
    \textbf{Wilko Heuten}\\
    \affaddr{Interactive Systems Group} \\
    \affaddr{OFFIS - Institute for IT} \\    
    \affaddr{Oldenburg, Germany} \\
    \email{wilko.heuten@offis.de} } \vfil \alignauthor{%
    \textbf{Andrii Matviienko}\\
    \affaddr{Media Informatics and Multimedia Systems} \\
    \affaddr{University of Oldenburg, Germany} \\    
    \email{andrii.matviienko@uni-oldenburg.de \smallskip} }\alignauthor{%
    \textbf{Susanne Boll}\\
    \affaddr{Media Informatics and Multimedia Systems} \\
    \affaddr{University of Oldenburg, Germany} \\    
    \email{susanne.boll@uni-oldenburg.de} } \vfil \alignauthor{%
    \textbf{Yannick Feld}\\
    \affaddr{Media Informatics and Multimedia Systems} \\
    \affaddr{University of Oldenburg, Germany} \\    
    \email{yannick.feld@uni-oldenburg.de} } \vfil  }

\def\plaintitle{VapeTracker: Tracking Vapor Consumption to Help E-cigarette Users Quit} \def\plainauthor{Abdallah El Ali, Andrii Matviienko, Yannick Feld,
  Wilko Heuten, Susanne Boll}
\def\plainkeywords{E-cigarettes; vaping; habits; health; tracking; sensors; behavior change technology; vapetracker; prototype}

\hypersetup{%
  pdftitle={\plaintitle}, pdfauthor={\plainauthor},
  pdfkeywords={\plainkeywords}, }

\begin{document}

\maketitle

\RaggedRight{} 

\begin{abstract}

Despite current controversy over e-cigarettes as a smoking cessation aid, we present early work based on a web survey ($N$=249) that shows that some e-cigarette users (46.2\%) want to quit altogether, and that behavioral feedback that can be tracked can fulfill that purpose. Based on our survey findings, we designed VapeTracker, an early prototype that can attach to any e-cigarette device to track vaping activity. We discuss our future research on vaping cessation, addressing how to improve our VapeTracker prototype, ambient feedback mechanisms, and the future inclusion of behavior change models to support quitting e-cigarettes.

\end{abstract}


  
\section{Introduction}

\begin{marginfigure}[-30pc]
    \centering
    \includegraphics[width=1\marginparwidth]{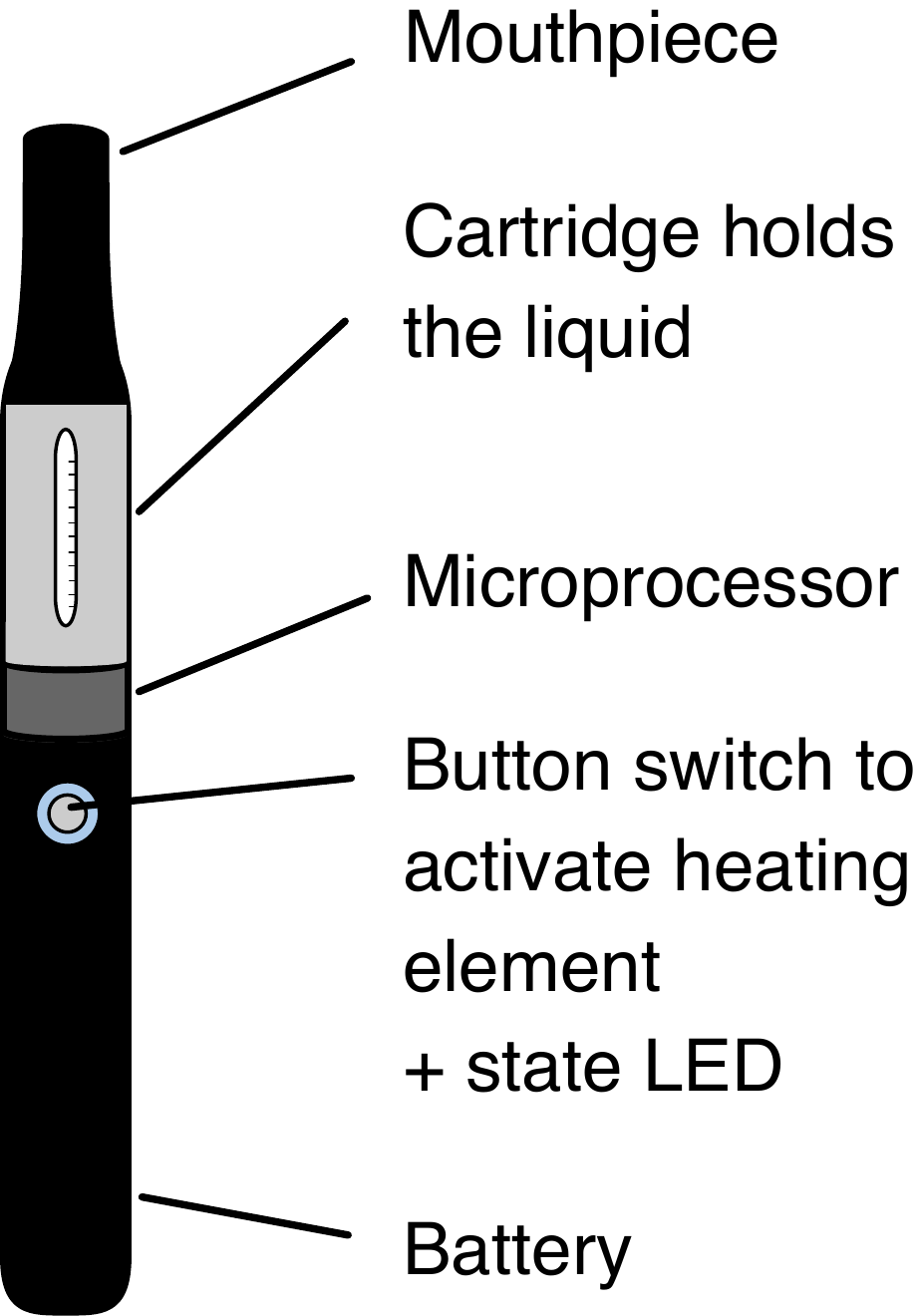}
    \caption{Parts of a second generation electronic cigarette (illustration based on KangerTech\textsuperscript{\textregistered} EVOD2).}
     \label{fig:ecig}
\end{marginfigure}

Since e-cigarettes were introduced in the market in 2004, they have been dubbed as a healthier alternative to smoking, and as an aid to help smokers quit \cite{Caponnetto2012}. Electronic cigarettes typically comprise a re-chargeable lithium ion battery, a battery powered atomizer which produces vapor by heating a solution of nicotine or a non-nicotine flavored solution, usually in propylene glycol or glycerine \cite{Britton2014}. This liquid is usually held in an often refillable cartridge in the device (Figure \ref{fig:ecig}). Drawing air through an e-cigarette (called `vaping') triggers the heating coils which creates vapor that simulates the feeling of smoking \cite{Caponnetto2012}. Since it produces vapor, no tobacco is burned and so is free of the many toxic chemicals present in tobacco smoke. Important to mention that third-generation e-cigarette devices (called `Mods', from modifications), consist of large-capacity lithium batteries with integrated circuits that allow vapers to change the voltage or power delivered to the atomizer which allows consumers to prepare their own setup of resistance and wick \cite{Farsalinos2014}. 

While current evidence suggests that e-cigarettes are indeed a healthier alternative to smoking, they are not without risks, especially since no standards are currently set \cite{Farsalinos2014,Palazzolo2013,Protano2015}. According to a recent 2015 report, there are currently 2.6 million adults in Great Britain using electronic cigarettes, of which approximately 1.1 million are ex-smokers\footnote{Source: \href{http://www.ash.org.uk/files/documents/ASH_891.pdf}{http://www.ash.org.uk/files/documents/ASH\_891.pdf} ; last retrieved: 11.1.2016}. Other 2015 statistics estimates that the total number of e-cigarette smokers in the United States is approximately 2.5 million\footnote{Source: UBS, Wells Fargo, Tobacco Vapor Electronic Cigarette Association; \href{http://www.statisticbrain.com/electronic-cigarette-statistics/}{http://www.statisticbrain.com/electronic-cigarette-statistics/} ; last retrieved: 11.1.2016}. It is a fact that this phenomenon of e-cigarette usage has increased globally \cite{Rom2015}, as additionally shown by Google trends of search term popularity over recent years\footnote{\href{https://www.google.com/trends/explore\#q=vaping}{https://www.google.com/trends/explore\#q=vaping} ; last retrieved: 11.1.2016 \bigskip }.

\section{Motivation \& Research Questions}
While several smoking cessation apps are available on various smart phone platforms \cite{Abroms2013}, HCI research studies on smoking cessation have been very few \cite{Paay2015b}. Given the rapid adoption of e-cigarettes, their use for smoking cessation, and their form, they provide a timely opportunity for embedding these electronic cigarettes with sensors to further raise user awareness about consumption, and facilitate behavior change for people who want to quit smoking and vaping altogether. In this paper, we take the first steps to address the role of sensor-based technology in facilitating vaping cessation, how it can be used together with these e-cigarette devices, and sketch out a research agenda to address this timely issue. Importantly, we also test our core assumption that some people would like to quit vaping altogether. 

To this end, we aim to address the following research questions: what information related to vaping activities should be tracked in order to aid in vaping cessation? How can we design a tracking device that can be used with e-cigarettes that fulfils such a goal? The rest of the paper will address related work on smoking cessation and behavior change technology, provide problem validation by using results of an online survey, present our early VapeTracker prototype, and sketch out future research areas.

\section{Related Work}

\subsection{Commercial Devices}

Closely related to our work are the Vapio\footnote{\href{http://www.vap.io/}{http://www.vap.io/} ; last retrieved: 11.1.2016} smart vaping e-cigarette device and Vaporcade\footnote{\href{https://vaporcade.com/}{https://vaporcade.com/} ; last retrieved: 11.1.2016} Jupiter\textsuperscript{TM} smartphone vaping device, where both allow tracking of users' everyday vaping activity and visualizing them as a graph, and Vapio can be used to track the user's location and allow syncing with the user's social network. While both relate to our work, their focus is not to research the role of behavior change technology and identify suitable factors for stopping e-cigarette usage. 

\subsection{Health Behavior Change \& Tracking}

Hekler et al. \cite{Hekler2013} classified four behavioral theories - meta-models, conceptual frameworks, constructs, and empirical findings. They identified three uses of behavioral theory in HCI: (1) inform the design of technical systems, (2) to guide evaluation strategies, and (3) to define target users. Klasnja et al. \cite{Klasnja2011} addressed the difficulties with evaluating behavioral change and that demonstrating behavior change is often not feasible as well as unnecessary for a meaningful contribution to HCI research, especially during early design stages or when evaluating novel technologies. Rather, HCI contributions should focus on efficacy evaluations tailored to specific intervention strategies (e.g., self-monitoring, conditioning). Ananthanarayan et al. \cite{Ananthanarayan2014} adopted a user-centric approach to behavior change, and studied how users can themselves craft personal health visualizations. They tested a wearable device based on participants' design choices that used an electronic cherry blossom tree as a visualization of physical activity, where the brighter the LED the more time was spent outside and found it to be an effective health behavior visualization tool. 

Shiffman \cite{Shiffman2005} highlighted the importance of dynamic changes in background conditions when testing Ecological Momentary Assessment (EMA) methods and in immediate states as important influences on smoking lapses and relapse, and raises the importance of considering situation interactions as key to cessation. The transtheoretical model (TTM) of behavior change \cite{Prochaska2008} assesses an individual's readiness to act on a new healthier behavior, and provides strategies, or processes of change to guide the individual through the stages of change to Action and Maintenance. However, \cite{Aveyard2009} found no evidence that the TTM-based intervention was more effective for participants in pre-contemplation or contemplation than for participants in preparation. 

Fogg's Behavior Model (FBM) asserts that for a person performing a target behavior, they require sufficient motivation, the ability to perform the target behavior as well as a trigger. Unless these three factors occur simultaneously, the target behavior will not occur \cite{Fogg2009}. Fenicio and Calvary \cite{Fenicio2015} evaluated the FBM triggers in their CRegrette system, testing standard smoking tools, a smartphone app, and a micro-controller-based ambient device, and found that behavioral statistics combined with ambient feedback are more effective than notifications, mirroring and self-monitoring approaches. Finally, Choe et al. \cite{Choe2014} studied how people use self-tracking technologies (through analyzing video recordings of Quantified Self Meetup talks), and highlight common pitfalls to self-tracking, including tracking too many things, not tracking triggers and context, and having insufficient scientific rigor in measurement and analysis of behavior.

\subsection{Smoking Cessation Technology}

Hoeppner et al. \cite{Hoeppner2015} found that publicly available smartphone smoking cessation apps are not particularly `smart', in that they commonly fall short of providing tailored feedback, despite users' explicit preference for such features. Paay et al. \cite{Paay2015} present the design and evaluation of a smartphone app called QuittyLink, designed to help smokers reduce or stop smoking. Their approach, which combined self-tracking of smoking activities and personal counselling, showed that both the personal counselling and the ability to visualize and reflect on self-tracked smoking behaviors helped participants form strategies to improve their ability to quit. Scholl et al. \cite{Scholl2013} present the UbiLighter prototypes, which can capture and record instances when the user smokes. Deploying their prototypes with 11 participants over several weeks, they found that smokers are generally unaware of their daily smoking patterns, and tend to overestimate their consumption. 

Furthermore, Ploderer et al. \cite{Ploderer2012} showed that smokers and recent ex-smokers are doubtful about their behavior change as well as about collecting personal information through technology and sharing it. Other work includes machine learning approaches to detect smoking activities, using RF proximity sensors \cite{Lopez-Meyer2013}, wrist-worn accelerometer data \cite{Tang2014}, or respiration measurements in the mPuff system \cite{Ali2012}.

\section{Web Survey}

To validate our assumption that people would like to quit smoking and vaping altogether, and to gain insight about e-cigarette user profiles and their information needs, we ran a web survey in Dec., 2015 for 1 week. It was distributed via online vaping forums\footnote{Spanning English-speaking users from across the world.}, and authors' social networks. 253 respondents filled the survey, however 4 were excluded due to noisy data. For the final analysis, total of 249 respondents (223 male\footnote{We acknowledge that our sample has a skewed distribution regarding gender since we recruited from vaping forums, however previous work has shown a current male dominance in usage of e-cigarettes (cf., 70\% males from sample of 1302 respondents in the survey by Dawkin et al. \cite{Dawkins2013}).} aged between 15 and 67 ($\bar{x}$ = 31.3, \textit{$s$} = 11) were included. Of these, 84.3\% are ex-smokers, 84.7\% used a third generation Mod device, 92.3\% stated they vape both in- and outdoors. 46.2\% stated they are vaping as a means to quit both smoking and vaping altogether, 28.1\% stated they do not intend to quit, and the remainder were unsure whether or not they want to quit (23.7\%).

\begin{marginfigure}[-18pc]
    \centering
            
    \begin{subfigure}[a]{\textwidth}\hspace{-5pt}

        \includegraphics[width=\textwidth]{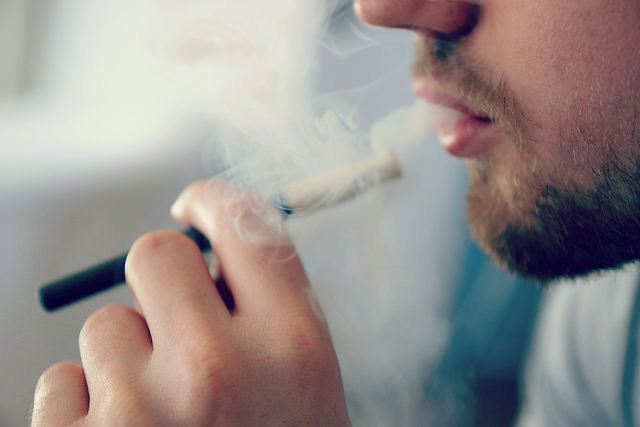}
    \caption{Grasp A}
	    \medskip
    \end{subfigure}

    \begin{subfigure}[b]{\textwidth}

        \includegraphics[width=\textwidth]{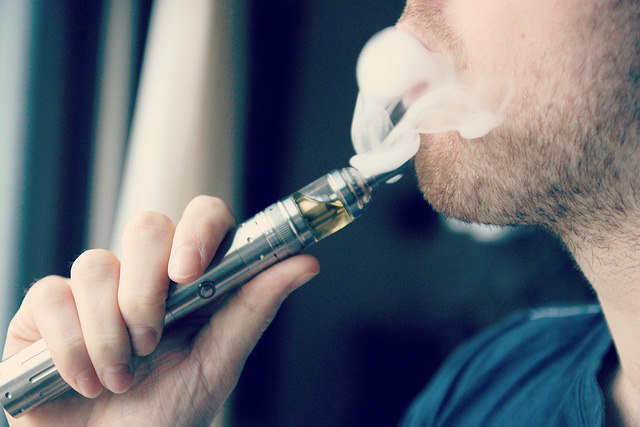}
            \caption{Grasp B}
	\medskip

    \end{subfigure}

    \begin{subfigure}[b]{\textwidth}

        \includegraphics[width=\textwidth]{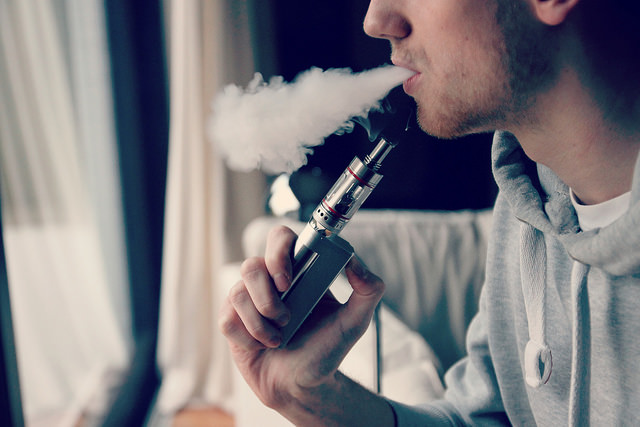}
            \caption{Grasp C}

    \end{subfigure}
    \caption{Different surveyed grasps of e-cigarette devices\protect\footnotemark. Image (a) shows typical grasping of first-generation e-cigarettes, image (b) shows grasping of second-generation e-cigarettes, and image (c) shows grasping of a third-generation Mod device.}\label{fig:vapegrasp}
\end{marginfigure}

\footnotetext{Images reused with permission from a Flickr\textsuperscript{\textregistered} Creative Commons attribution license. Credit: Vaping360.com}

As for vaping behavior related to refilling nicotine liquid, 9.2\% used less than 3mg nicotine bottles, 44.1\% used 3-6mg, 26\% used 6-9mg, 13.7\% greater than 9mg, and 6.8\% did not specify in detail. For how many refills\footnote{For those that do not use a dripping mechanism, which constantly refills the e-cigarette.} per day, 24.9\% refilled once a day, 14.1\% twice a day, 8\% three times a day, 13.7\% more than 3 times per day, 11.7\% used dripping, and 27.7\% used other variations\footnote{Including unspecified counts per day or once every few days.}.

\subsection{Results}

Two 6-point Likert items ($\alpha$ = 0.67) asked respondents about whether they were aware of how much they vaped during the last week and whether they felt they lost track of how much they vaped on a given day. Items showed a medium-sized correlation ($r_s$ = 0.53, $p$ < 0.001)\footnote{Spearman's rank correlation coefficient was used due to the ordinal nature of Likert items.}, and were thereafter binned into agree/disagree responses. Only 16\% of respondents said they were not aware of how much they vaped in the last week, and only 20\% stated they lost track of how much they vaped on a given day. These results are surprising, as it is very difficult to count exactly how much they vape per day, especially since users are not limited to where they vape and how much liquid they use up\footnote{Further, anecdotal user reports on e-cigarette forums report nicotine overdosing, however little medical research to date addresses this \cite{Farsalinos2014b}.}. Furthermore, we did not find a significant effect between quitting vaping intention (yes, no, unsure) and respondent awareness of how much they vaped in the last week ($\chi^2$ = 0.669, $p$ = 0.72) nor whether respondents lost track of how much they vape ($\chi^2$ = 1.045, $p$ = 0.59). 

When we asked which e-cigarette device grasping behavior (Figure \ref{fig:vapegrasp}) most closely resembles how a respondent holds her device, grasp C received the highest counts (220), followed by B (55) and A (27). This is not surprising given that most of our sample used third generation devices. This provided us with hints that despite the seemingly unergonomic factor of these larger Mod devices, people are willing to carry them. This provides ample opportunity for inclusion of micro-controllers and sensors for tracking, where below we describe our early VapeTracker prototype.

\begin{center}

\begin{table}
\small
\begin{tabular}{p{6cm}P{1.4cm}}
  \hline
  \textbf{Feedback Type} & \textbf{Frequency} \\ \hline
  Vape counts per day/week/month & \textbf{72} \\
  Vaping session counts per day/week/month & 52 \\
  Day/week/month with highest vape count & 23 \\
  Comparison to number of smoked cigarettes & \textbf{145} \\
  Locations of highest vape activity & 39 \\
  Person(s) who I vape with & 26 \\
  Comparison of vaping activity to other vapers & \textbf{85} \\
  Such information would not help quitting	& \textbf{66}\\
  Other & 26	\\
  \hline
\end{tabular}
\caption{Feedback type to support vaping cessation and respondent counts. Bold values show high frequencies.} 
\label{table:info}
\end{table}
\end{center} 
\vspace{-2pc}

\begin{marginfigure}[2pc]
  \begin{minipage}{\marginparwidth}
    \centering
    \includegraphics[width=0.9\marginparwidth]{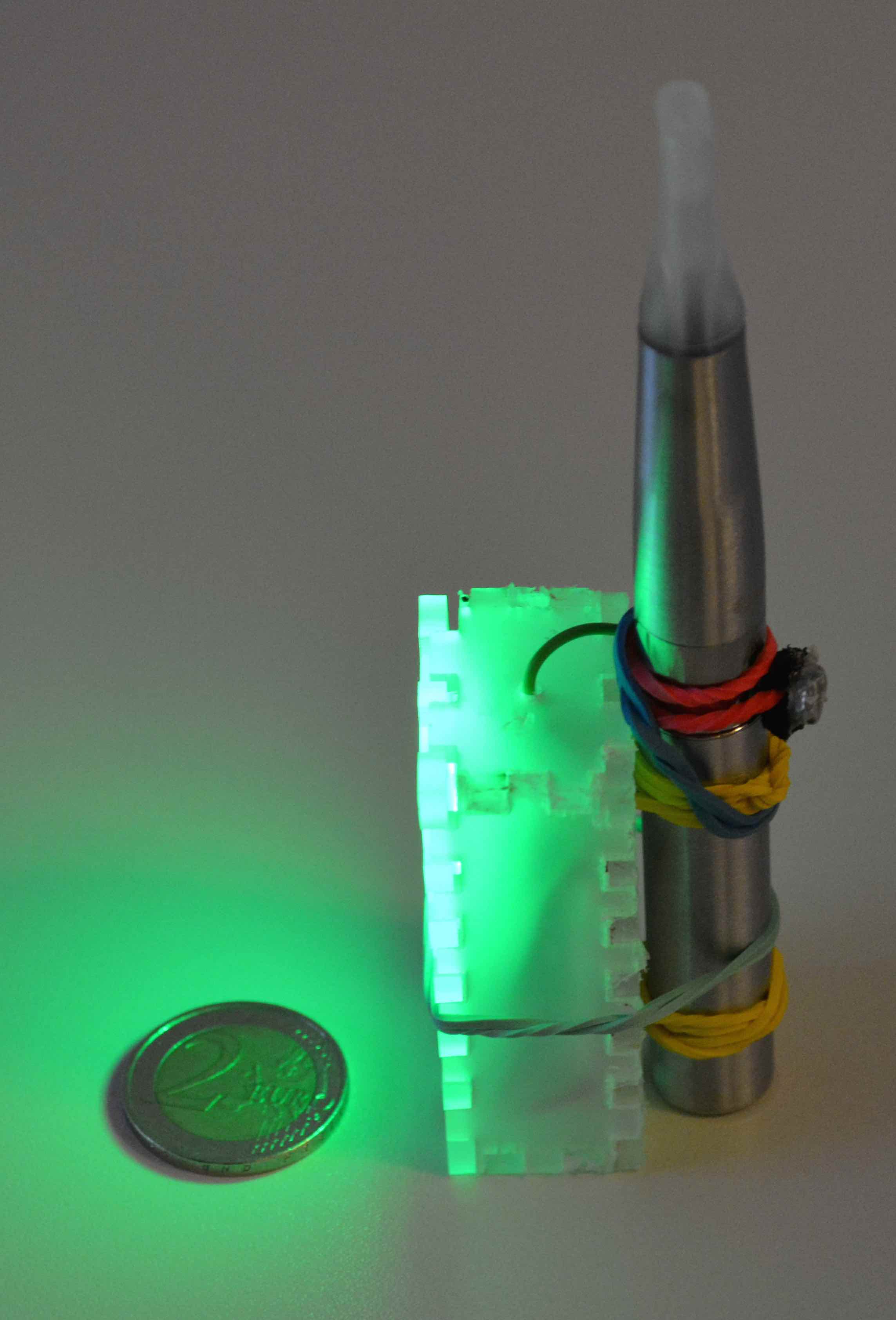}
    \caption{Our early VapeTracker prototype (attached to the Joyetech\textsuperscript{\textregistered} eGo-C e-cigarette) emitting green ambient light, placed next to a \euro{}2 coin for size comparison.}~\label{fig:vapetracker_prototype}
  \end{minipage}
\end{marginfigure}

Finally, when we asked about what feedback type would respondents find most helpful if they were attempting to quit vaping (provided through checkboxes), the highest count frequencies were for: Comparison to number of smoked cigarettes, vape counts per day/week/month, and comparison of vaping activity to other vapers. These results are summarized in Table \ref{table:info}. While comparison to cigarettes smoked seems to be the most promising, this is also the most problematic feedback type due to the high variability in a direct translation from vaping to smoking. In order to obtain a meaningful direct mapping, one would have to estimate the nicotine blood plasma levels at a given point, accounting for other variables such as puff duration, puff frequency, nicotine strength, type of vaporizer, and individual nicotine absorption levels. Therefore, in our early prototype we have kept our focus on tracking activities that do not require medical instrumentation.

\section{Early VapeTracker Prototype}

Based on our results that some people would like to quit vaping altogether, that vape counts per day/week/month matter to users who want to quit, and on Fenicio and Calvary's \cite{Fenicio2015} findings that showed the effectiveness of ambient notifications and statistics, we built VapeTracker. VapeTracker is an early prototype that can be attached to any e-cigarette for tracking the number of puffs and puff duration, and in the future visualize this activity to help users become more aware of their vape activities, and eventually quit.

Our first prototype (see Figure \ref{fig:vapetracker_prototype}) is based on a modified LightBlue\textsuperscript{\textregistered} Bean\footnote{\href{https://punchthrough.com/bean}{https://punchthrough.com/bean} ; last retrieved: 11.1.2016} programmable board, which contains an ATmega 328p microcontroler, 3-axis accelerometer and Bluetooth LE Peripheral (CC2540 BLE radio). In addition to the Bean, we used a 3.7V Lithium Polymer (LiPo) battery, a micro-USB LiPo charger, and three Adafruit\textsuperscript{\textregistered} RGB Neopixels. The coin cell slot was removed from the Bean in order to make space of the LiPo battery, which lasts longer. 

We used the Bean for ergonomic reasons due to its small size (45.5 x 20.3 x 8.38 mm) and integrated Bluetooth and accelerometer. We used a capacitative button placed on the e-cigarette power button that has a wired connection to the Bean. All button presses and durations are saved in internal board memory (EEPROM), which is capable of saving 1 Kb of data for a period of 2.86 days if unused\footnote{Given a delay time of 10ms between button presses and parameter $k$ in our compression algorithm that gives higher weight to cases when the button is unpressed.} and 0.8 days for continuous vape tracking. Button press false positives are handled by currently logging only press durations of >500 ms. For this prototype, we initialize the board using an Arduino program, and use another program for logging data. In this flow, the Bean executes a program (via Bluetooth) designed to read the data, saves and processes the values internally, and determines the color of the LED. Currently we do not have an integrated real-time clock (RTC), thus during the logging phase we request timestamps over a serial port. Currently we do not log accelerometer data, however later we can investigate quantitatively vaping gestures, or as a means for a user to choose between on-device ambient visualizations. 

\begin{marginfigure}[-1pc]\hspace{-2pc}
  \begin{minipage}{\marginparwidth}
    \centering
    \includegraphics[width=1.2\marginparwidth]{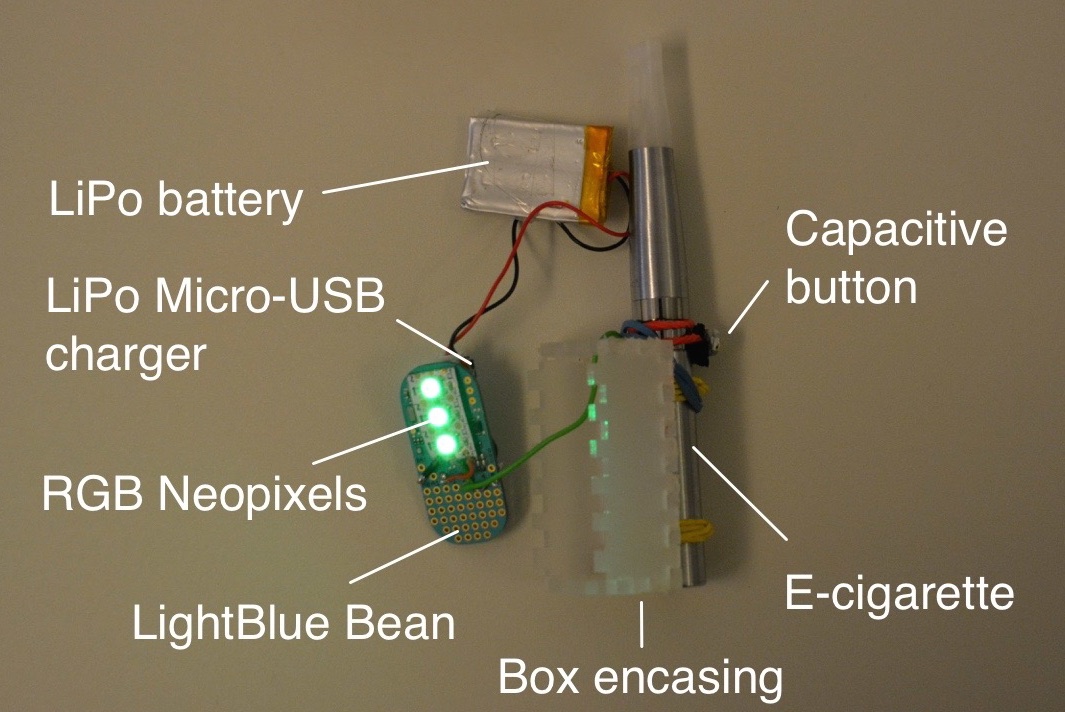}
    \caption{Hardware components of our early VapeTracker prototype.}~\label{fig:vapetracker_parts}
  \end{minipage}
\end{marginfigure}

All hardware components (shown in Figure \ref{fig:vapetracker_parts}) are enclosed into a laser-cut box shaped as a rectangular parallelepiped made of diffused acrylic glass. We used a semi-transparent glass to ensure the visibility of ambient light feedback, while simultaneously ensuring hardware components are hidden from the user. The size of this box (58 x 32 x 22 mm) is smaller than currently available Mod vape devices (where all fit within the average palm width\footnote{\href{http://www.theaveragebody.com/average_hand_size.php}{http://www.theaveragebody.com/average\_hand\_size.php} ; last retrieved: 11.1.2016} of males (84 mm) and females (74 mm)). However, for this early prototype the box size was chosen solely based on the space needed for components to ensure robust, everyday usage.

\section{Next Steps \& Research Agenda}

Our immediate next steps are to iterate over our first prototype, by using Bluetooth as the communication channel between the VapeTracker prototype and a smartphone, which would handle both timestamp fetching and data storage.This would additionally allow us to experiment with tracking frequent vaping locations, despite that vaping location did not receive a high respondent count. Once a testable prototype is in place, we plan to investigate the form factor of VapeTracker, ensuring ergonomics, usability, and social acceptance. Further on, we intend on gathering further insight from users on the role of ambient on-device feedback (using LEDs): What should the status indicators represent (e.g., vape consumption as a progress bar)? How should the user switch on and between visualizations (e.g., shaking gestures)? We plan on adopting a user-centric approach similar to the work by Ananthanarayan et al. \cite{Ananthanarayan2014} where users crafted their visualizations. Following that, we need to consider a screen-based vaping activity dashboard: which device should it be shown on (smartphone, watch, desktop, or a combination)? What are the most useful behavioral statistics and indicators that e-cigarette users find? 

These considerations cannot be separated from psychological models of behavioral change, therefore in parallel we need to investigate more thoroughly which model is best suited for vapers who want to quit, and the importance of immediate relapse precipitants (cf., \cite{Shiffman2005}). At that stage, our plan is to build several VapeTracker prototypes and deploy to users for getting longitudinal data (order of months) as is done in smoking cessation HCI research \cite{Paay2015,Scholl2013}. Finally, a long-term goal of our work is to eventually acquire enough vaping data that machine learning models can be trained, in order to predict when a user would vape, how to personalize feedback and provide smart user-aware notifications. 

In summary, in spite of current controversy over e-cigarettes as a smoking cessation aid, our early work has validated the assumption that some e-cigarette users want to quit altogether, and that many users use Mod devices, which allow inclusion of sensors for behavioral tracking and subsequent intervention. From this, we have designed our early VapeTracker prototype to track vaping activities. Through our ongoing work, we hope to have set the stage for further novel technology research concerning this timely issue.

\section{Conflict of Interest}
Authors certify that they have no affiliations with or involvement in any organization or entity with any financial or non-financial interest for promoting or demoting e-cigarettes.
\balance{} 

\bibliographystyle{SIGCHI-Reference-Format}
\bibliography{chi16_vapetracker}

\end{document}